\documentclass[a4paper,12pt]{article}
\usepackage{amssymb,amsmath}
\usepackage{graphics,amsfonts,epsfig}

\title{Hyperbolic calorons, monopoles, and instantons}
\author{Derek Harland\footnote{email address: d.g.harland@durham.ac.uk}
  \bigskip
  \\Deparment of Mathematical Sciences,
  \\Durham University,
  \\DH1 3LE}
\date{}

\begin{document}

\maketitle
\begin{abstract}
We construct families of $SO(3)$-symmetric charge 1 instantons and calorons on the space $\mathbb{H}^3 \times \mathbb{R}$.  We show how the calorons include instantons and hyperbolic monopoles as limiting cases.  We show how Euclidean calorons are the flat space limit of this family.
\end{abstract}

\section{Introduction}
\label{introduction}
Calorons are instantons on $\mathbb{R}^3\times S^1$.  It has been known for some time that they are related both to instantons on $\mathbb{R}^4$ and BPS monopoles on $\mathbb{R}^3$.  For example, the large period limits of calorons are normally instantons \cite{hs}, while the large scale limit of a charge 1 caloron is a charge 1 monopole \cite{rossi}.

It is interesting to study monopoles on hyperbolic space $\mathbb{H}^3$, \cite{atiyah} \cite{nash} \cite{chakrabarti85}.  These ``hyperbolic monopoles'' are related to their Euclidean counterparts, because one can recover Euclidean monopoles in the limit where the curvature of hyperbolic space tends to zero \cite{jn97}.  Hyperbolic monopoles where first constructed by means of a conformal equivalence \cite{atiyah} \cite{chakrabarti85}: since $\mathbb{R}^4$ is conformally equivalent $\mathbb{H}^3\times S^1$, hyperbolic monopoles can be obtained from instantons invariant under an action of $U(1)$.  Hyperbolic monopoles constructed in this way have the property that the asymptotic norm of their Higgs field is always an integer, and are sometimes called ``integral'' for this reason.  Non-integral hyperbolic monopoles were constructed later by different means \cite{nash}.

The conformal equivalence used to construct integral hyperbolic monopoles has also been used to construct hyperbolic calorons, that is, instantons on $\mathbb{H}^3\times S^1$ \cite{gm}.  Hyperbolic calorons constructed in this way have the property that their period (the circumference of $S^1$) is proportional to the radius of curvature of the hyperbolic space.  This property is the analogue of the integral property which arises when hyperbolic monopoles are constructed from Euclidean instantons.  The purpose of this article is to show the existence of hyperbolic calorons which do not have this integral property.  We will give explicit examples of charge 1 calorons with arbitrary period on hyperbolic spaces with arbitrary curvature.  We will show that these are related to Euclidean calorons and hyperbolic monopoles in the ways that one might expect; we will also show that they have a well-defined large period limit, which is an instanton on $\mathbb{H}^3\times\mathbb{R}$.

We will now briefly outline the content of the remainder of this article.  In section \ref{definitions}, we will give precise definitions for hyperbolic instantons and calorons.  In section \ref{construction}, we will outline two simple methods that can be used to construct them.  In section \ref{examples} we will give explicit examples of hyperbolic calorons and instantons, and in section \ref{properties} we will explore some of their properties.  We will conclude with some brief remarks in section \ref{conclusion}.

\section{Hyperbolic calorons and instantons}
\label{definitions}

Let $(x^1,x^2,x^3)$ be coordinates on the hyperbolic ball $\mathbb{H}^3$ and let 
\[ R=\sqrt{(x^1)^2 + (x^2)^2 + (x^3)^2} \]
be the radial coordinate, with $0\leq R < S$ for some fixed $S$ ($S$ will determine the scalar curvature of $\mathbb{H}^3$).  Let $\tau=x^0$ be a coordinate on $S^1$, with period $\beta$.  The metric on $\mathbb{H}^3\times S^1$ is
\[ ds_H^2 = d\tau^2 + \Lambda^2 (dR^2 + R^2 d\Omega^2), \]
where $\Lambda:=(1-(R/S)^2)^{-1}$ and $d\Omega^2$ represents the metric on the 2-sphere.  It will also be convenient to introduce a coordinate $\mu = (S/2) \mbox{arctanh} (R/S) $ and a complex coordinate $z=\mu+i\tau$.  In terms of $\mu$ and $\tau$, the metric is
\[ ds_H^2 = d\tau^2 + d\mu^2 + \Xi^2 d\Omega^2, \]
with $\Xi := (S/2)\sinh\left(2\mu/S\right)$.

Let $(y^0,y^1,y^2,y^3)$ be standard coordinates on $\mathbb{R}^4$, and let $t=y^0$ and 
\[r=\sqrt{(y^1)^2 + (y^2)^2 + (y^3)^2 }. \]
The metric on $\mathbb{R}^4$ is
\[ ds_E^2 = dt^2 + dr^2 + r^2 d\Omega^2. \]
We let $Z=r+it$ be a complex coordinate.

When $\beta=S\pi$, $\mathbb{H}^3\times S^1$ is conformally equivalent to $\mathbb{R}^4 \backslash \mathbb{R}^2$.  We define a map from $\mathbb{H}^3\times S^1$ to $\mathbb{R}^4$ by the equation $Z=\tanh(z/S)$.  Then it is simple to show that
\[ ds_H^2 = \xi^2 ds_E^2, \]
with $\xi = (S/2) ( \cosh(2\mu/S) + \cos(2\tau/S))$.

Gauge fields on $M = \mathbb{H}^3\times S^1$ or $\mathbb{H}^3\times\mathbb{R}$ will be denoted $A = A_\alpha dx^\alpha$, with $A_\alpha$ traceless anti-hermitian matrix-valued functions.  The field strength tensor of a gauge field is $F = (1/2) F_{\alpha\nu} dx^\alpha \wedge dx^\nu$, where $F_{\alpha\nu} = \partial_\alpha A_\nu - \partial_\nu A_\alpha + [A_\alpha,A_\nu]$.  The action of a gauge field on $M$ is
\[ S = -\frac{1}{8} \int_{M} \mbox{Tr} (F_{\alpha\nu} F^{\alpha\nu}) \, \Xi^2 \, d\tau \, d\mu \, d\Omega \]
($d\Omega$ represents the volume form on the 2-sphere).

Hyperbolic calorons are defined to be gauge fields on $\mathbb{H}^3\times S^1$, whose curvature satisfies the self-dual equations,
\begin{equation}
\label{sd equation}
F_{0i} = \frac{1}{2\Lambda} \epsilon_{ijk} F_{jk}
\end{equation}
and whose action is finite, with $\mbox{Tr}(F_{\alpha\nu}F^{\alpha\nu})\rightarrow0$ as $\mu\rightarrow\infty$.  Similarly, hyperbolic instantons are self-dual gauge fields on $\mathbb{H}^3\times\mathbb{R}$ with finite action and with $\mbox{Tr}(F_{\alpha\nu}F^{\alpha\nu})\rightarrow0$ as $\mu^2+\tau^2 \rightarrow\infty$.

Finite action gauge fields on $M=\mathbb{H}^3\times S^1$ or $\mathbb{H}^3\times\mathbb{R}$ have a charge,
\[ W = - \frac{1}{8\pi^2} \int_{M} \mbox{Tr} (F\wedge F), \]
The action is bounded from below by the charge, $S\geq 2\pi^2 |W|$.  Instantons and calorons attain this bound and have $W>0$, since they solve (\ref{sd equation}).  For instantons, $W$ is an integer.  This follows from usual argument: the gauge field extends to the manifold $S^3$ at infinity, and since the curvature is zero there, the gauge field is zero up to a gauge transformation $g:S^3\rightarrow SU(2)$.  The map $g$ has integer degree, which is computed by $W$.  The situation for calorons is a bit more complicated.  A caloron is characterised by two integer topological charges, called the instanton charge (or Pontryagin index) and the magnetic charge, and a boundary condition, called the holonomy \cite{gpy}.  The integral $W$ is equal to a combination of these three quantities, and is not necessarily an integer.  However, the calorons we consider here will have zero magnetic charge and trivial holonomy, and in this case $W$ will be an integer.

\section{Construction}
\label{construction}
As we have seen, the metric $ds_H^2$ on $\mathbb{H}^3\times S^1$ is locally conformally equivalent to the metric $ds_E^2$ on $\mathbb{R}^4$.  It is well known that the action of the Hodge star operator on 2-forms on $\mathbb{R}^4$ is invariant under conformal rescalings of the metric.  Therefore the the self-dual equations on $\mathbb{H}^3\times S^1$ are locally the same as the self-dual equations on $\mathbb{R}^4$.  There are a number of methods available for constructing self-dual gauge fields on $\mathbb{R}^4$, and all of these can be applied directly on $\mathbb{H}^3\times S^1$.  In this section we will show how two well-known constructions for Euclidean instantons can be applied to give locally self-dual gauge fields on $\mathbb{H}^3\times S^1$.  We will show later that these methods can in fact be used to generate finite action gauge fields, that is, hyperbolic calorons (and hyperbolic instantons).  

\subsection{Harmonic function ansatz}
\label{harmonic function ansatz}
The first ansatz we shall consider is the harmonic function ansatz, due to Corrigan, Fairlie, and t'Hooft \cite{cf}.  On Euclidean space, one supposes that the gauge field is written in the form,
\begin{eqnarray}
\label{euclidean hfa}
A_0 &=& \partial_j \ln\,\varphi (\sigma^j/2i) \\
\nonumber A_j &=& (-\partial_0 \ln\,\varphi \, \delta_{jl} + \epsilon_{jlk}\partial_k\ln\,\varphi)(\sigma^l/2i)
\end{eqnarray}
for some function $\varphi$.  Then the gauge field will be self dual when $\varphi$ satisfies the Laplace equation,
\[ \Box_E \varphi = 0, \]
where $\Box_E = (\partial/\partial y^\alpha)^2$ is the (Euclidean) Laplacian.

To use this ansatz on the metric $ds_H^2$, we let $\rho = \varphi/\xi$ and change coordinates in the above expressions.  The resulting ansatz for the gauge field is
\begin{eqnarray}
\label{hyperbolic hfa1}
A_0 &=& \frac{1}{\Lambda} \partial_j \ln \rho \left( \frac{\sigma^j}{2i} \right) \\
\label{hyperbolic hfa2}
A_j &=& \left( -\Lambda\partial_0 \ln \rho \, \delta_{jl} + \epsilon_{jlk} \left( \partial_k \ln\rho + \frac{2\Lambda}{S^2} x^k \right) \right) \left( \frac{\sigma^l}{2i} \right).
\end{eqnarray}
This will be self-dual if $\rho$ satisfies the equation,
\begin{equation}
\label{DE for rho}
\Box_H \rho = - \frac{4}{S^2} \rho
\end{equation}
where $\Box_H$ is the Laplace-Beltrami operator for the metric $ds_H^2$,
\[ \Box_H = \left(\frac{\partial}{\partial\tau}\right)^2 + \frac{1}{\Lambda^2} \left(\frac{\partial}{\partial x^{i}}\right)^2 + \frac{2}{\Lambda S^2} x^i \frac{\partial}{\partial x^i}. \]

\subsection{$SO(3)$-symmetric ansatz}
\label{SO(3)-symmetric ansatz}
The second ansatz we shall consider is due to Witten \cite{witten}.  Witten simplified the problem of finding Euclidean instantons by restricting attention to gauge fields invariant under an action of $SO(3)$; he was able to solve the reduced self-duality equations exactly, yielding a large family of instantons.

Interestingly, the $SO(3)$-invariant instantons turned out to be related to a two-dimensional vortices.  The components of the $SO(3)$-invariant gauge were interpreted as a $U(1)$ gauge field and a 2-component Higgs field, and the action of the 4-dimensional gauge field was identical to that of a vortex model on 2-dimensional hyperbolic space.  Configurations in the vortex model had a topological charge, and this was equal to the topological charge of the corresponding 4-dimensional gauge theory.

Witten's method is also applicable to self-dual gauge fields on $\mathbb{H}^3\times S^1$ and $\mathbb{H}^3\times \mathbb{R}$, because the $SO(3)$ action pulls back to these manifolds.  We will see that the dimensionally-reduced equations of motion are solvable using an adaptation of Witten's method.  As in the Euclidean case, hyperbolic instantons will be related to a 2-dimensional vortex model, but this time the underlying 2-manifold will not be hyperbolic space.  In this section, our notation will closely follow that of \cite{landweber}

We make the following $SO(3)$-symmetric ansatz for a gauge field on $\mathbb{H}^3\times S^1$ or $\mathbb{H}^3\times\mathbb{R}$:
\begin{equation}
\label{symmetric gf}
A = -\frac{1}{2} (Qa + \phi_1 dQ + (\phi_2+1) QdQ ),
\end{equation}
where $Q=x^a\sigma^a/R$, $a=a_\mu d\mu + a_\tau d\tau$, and $\phi_1$, $\phi_2$, $a_\mu$, and $a_\tau$ are real functions of $\mu$ and $\tau$.  

We let $\phi = \phi_1-i\phi_2$ be a Higgs field, and let $D_\mu \phi=\partial_\mu\phi+ia_\mu\phi$ and $D_\tau \phi=\partial_\tau\phi+ia_\tau\phi$ denote the components of its covariant derivative with respect to the gauge field $a$.  We will also write $D\phi = D_\mu\phi d\mu + D_\tau \phi d\tau$ and $D_{\bar{z}} = (D_\mu + iD_\tau)/2$.  Then the self-dual equations (\ref{sd equation}) for $A$ can be succinctly written,
\begin{eqnarray}
\label{vortex eom4}
D_{\bar{z}} \phi &=& 0 \\
\label{vortex eom3}
\Xi^2 (\partial_\tau a_\mu - \partial_\mu a_\tau) &=& 1-|\phi|^2.
\end{eqnarray}

The action of the gauge field $A$ is equal to
\begin{equation}
\label{vortex action}
S = \frac{\pi}{2} \int \left( \Xi^2 (\partial_\mu a_\tau - \partial_\tau a_\mu)^2 + \left(\frac{1-|\phi|^2}{\Xi}\right)^2 + 2|D_\mu\phi|^2 + 2|D_\tau\phi|^2 \right) d\tau d\mu.
\end{equation}
This is the action of a vortex model on the 2-manifold $\mathcal{M}$ coordinatised by $\mu$ and $\tau$, with metric
\[ ds_h^2 = \Xi^{-2}(d\mu^2+d\tau^2). \]
In the case of hyperbolic instantons ($\beta=\infty$), the manifold $\mathcal{M}$ is a universal cover of the hyperbolic disc with a point removed.

Vortex configurations on $\mathcal{M}$ posess a topological charge, $k$, given by
\[ k = -\frac{1}{2\pi} \int da, \]
A standard Bogomolny argument shows that $2\pi^2k$ forms a lower bound for the action (\ref{vortex action}) within the set of vortex configurations with charge $k>0$; the equations (\ref{vortex eom4}), (\ref{vortex eom3}) are precisely the equations which guarantee the bound is attained.  It follows that $k=W$ for self-dual gauge fields.

We note that the ansatz (\ref{symmetric gf}) does not fix the gauge; one is free to make gauge transformations of the form $g=\exp(\lambda Q/2)$ for some real function $\lambda(\mu,\tau)$.  These gauge transformations correspond to $U(1)$ gauge transformations in the vortex model.

The solution of the self-dual equations (\ref{vortex eom4}), (\ref{vortex eom3}) is relatively simple.  One first needs to choose a meromorphic function $g(z)$ satisfying $|g|^2\leq1$, with equality when $\mu=0$, and a non-zero holomorphic function $h(z)$.  Then a self-dual gauge field is obtained from
\begin{eqnarray}
\label{phi}
\phi &=& e^\psi h \partial_z g \\
\label{a_mu}
a_\mu &=& -\partial_\tau \psi \\
\label{a_tau}
a_\tau &=& \partial_\mu \psi \\
\label{psi}
\psi &=& \ln(2\Xi) - \ln(1-|g|^2) - \ln |h|.
\end{eqnarray}
The function $h$ can be removed by gauge transformation ($h$ is included here for later convenience).  The derivation of these solutions is similar to that given by Witten in the Euclidean case.

\subsection{Equivalence}
\label{equivalence}
After the discoveries of the $SO(3)$-symmetric ansatz and the harmonic function ansatz for Euclidean instantons, Manton showed that the latter encompasses the former \cite{manton}.  More specifically, all self-dual gauge fields obtained via Witten's ansatz can also be expressed in the harmonic function form (\ref{euclidean hfa}).  A similar result holds for the two hyperbolic ans\"{a}tze discussed above.

Suppose that a hyperbolic gauge field is given by (\ref{symmetric gf}) and (\ref{phi})-(\ref{psi}) in terms of a meromorphic function $g$.  Then the gauge $h$ can be chosen so that the gauge field is in the form (\ref{hyperbolic hfa1}), (\ref{hyperbolic hfa2}), with
\begin{equation}
\label{rho from g}
\rho = \frac{1}{2\Xi} \left( \frac{1-|g|^2}{|1-g|^2} \right)
\end{equation}
satisfying (\ref{DE for rho}).  This result can be derived by following the method of Manton.

\section{Examples}
\label{examples}
We will now give examples of hyperbolic calorons and instantons, using the $SO(3)$-symmetric ansatz (\ref{symmetric gf}).

\subsection{Charge 1 instanton}
A charge 1 instanton is obtained from (\ref{phi})-(\ref{psi}) when
\[ g(z) = g_1(z) := \exp\left(-\frac{2z}{S}\right) \left( \frac{\lambda-2z/S}{\lambda+2z/S} \right), \]
where $\lambda>0$ is a real parameter.

To verify that this gauge field is an instanton, one must check that the action is finite.  This has been done directly, by computing all of the terms in the action (\ref{vortex action}) and verifying that their integrals converge.  The charge of the instanton was computed using methods from chapter II of \cite{jt}.

\subsection{Charge $k$ instanton}
Consider the function,
\[ g_k(z) := \exp\left( -\frac{2z}{S} \right) \prod_{i=1}^k \frac{\lambda_i-2z/S}{\bar{\lambda_i}+2z/S} \]
with $\lambda_i$ complex parameters with positive real part.  It is believed that when $g(z)=g_k(z)$ the gauge field is a charge $k$ instanton.  Certainly, $g_k(z)$ agrees with $g_1$ when $k=1$, and the Higgs field $\phi$ has $k$ zeros when $g=g_k$.  Normally the charge of a vortex configuration is equal to the number of zeros of the Higgs field.  Numerical results suggest that the action of the gauge field is finite and proportional to $k$, however, we have not yet proved this analytically.

\subsection{Charge 1 caloron}
The following function is a special case of $g_k$ with $k=\infty$: 
\[ g_\infty(z) := \exp \left( -\frac{2z}{S} \right) \lim_{k\rightarrow\infty}\prod_{j=-k}^{k} \frac{\lambda + 2j\beta i/S - 2z/S} {\lambda- 2j\beta i/S + 2z/S}, \]
where $\lambda$ and $\beta$ are positive real parameters.  The infinite product in $g_\infty(z)$ converges uniformly on any compact set, after singular terms have been removed, and the limit has the following closed form:
\[ \lim_{k\rightarrow\infty} \prod_{j=-k}^{k} \frac{\lambda + 2j\beta i/S - 2z/S} {\lambda- 2j\beta i/S + 2z/S} = \frac{ \sinh((z-S\lambda/2)\pi/\beta) }{ \sinh((z+S\lambda/2)\pi/\beta)}. \]
The gauge choice $h=\exp(2z/S)$ makes the gauge field obtained from $g_\infty$ explicitly periodic, with period $\beta$.  This gauge field is in fact a hyperbolic caloron with finite action and charge 1, as has been verified analytically.

We mentioned in the introduction that integral hyperbolic calorons (with $\beta=S\pi$) can be obtained directly from Euclidean instantons.  For example, a charge 1 instanton in the form (\ref{euclidean hfa}), with
\[ \varphi = 1 + \frac{\alpha^2}{r^2+t^2} \]
is also a charge 1 integral hyperbolic caloron.  One can also obtain charge 1 integral hyperbolic calorons from $g_\infty$ simply by setting $\beta=S\pi$.  In fact these two sets of integral hyperbolic caloron are identical; if one takes $\alpha^2=\tanh(\lambda/2)$ then the two calorons are related by a gauge transformation.  Therefore, the hyperbolic calorons obtained from $g_\infty$ are a sensible generalisation of the known integral charge 1 calorons.

\section{Properties}
\label{properties}
The non-integral charge 1 hyperbolic calorons we have constructed are determined by three parameters, $S$, $\beta$ and $\lambda$, which determine the curvature of the hyperbolic space, the period of the caloron, and the shape of the caloron.  By taking limits of these parameters we are able to relate the hyperbolic calorons to Euclidean calorons, hyperbolic instantons, and non-integral hyperbolic monopoles.

\subsection{Zero curvature limit}
When $S\rightarrow\infty$, the hyperbolic metric $ds_H^2$ converges to the Euclidean metric $ds_E^2$ if one makes the identifications $\tau=t$, $\mu=r$.  Charge 1 calorons were first constructed on Euclidean space by Harrington and Shepard \cite{hs} using the ansatz (\ref{euclidean hfa}), with
\[ \varphi = 1 + \frac{\pi\nu^2}{\beta r} \frac{ \sinh(2\pi r/\beta) } { \cosh(2\pi r/\beta)-\cos(2\pi t/\beta )} \]
Here $\nu>0$ is a real parameter and $\beta$ determines the period of the caloron.  We will show here that Euclidean 1-calorons are obtained from hyperbolic 1-calorons by taking a limit where $S\rightarrow\infty$.

The zeros of the Higgs field $\phi$ obtained from (\ref{phi}) when $g=g_\infty$ are located at the points $(\mu,\tau)$ determined by
\begin{eqnarray*}
\tau &=& n\beta \,, n\in\mathbb{Z} \\
\cosh\left(\frac{2\pi\mu}{\beta}\right) &=& \cosh\left(\frac{S\pi\lambda}{\beta}\right) + \frac{S\pi}{\beta} \sinh\left(\frac{S\pi\lambda}{\beta}\right).
\end{eqnarray*}
In order that these points remain finite as $S\rightarrow\infty$, we must also let $\lambda\rightarrow0$ such that $\lambda S^2$ remains finite.  In the gauge $h=(S/2)\exp(2z/S)$, the functions $\phi$, $a_\mu$, $a_\tau$ converge pointwise.  In fact, it can be shown that their limit describes a Euclidean 1-caloron, with
\[ \nu^2 = \lim_{S\rightarrow\infty} \frac{\lambda S^2}{2}. \]

\subsection{Instanton limit}
In the limit $\beta\rightarrow\infty$, the function $g_\infty(z)$ converges pointwise to $g_1(z)$, so one might expect the hyperbolic caloron gauge field to converge to a hyperbolic instanton gauge field in this limit.  This can be verified directly, using expressions derived from (\ref{phi})-(\ref{psi}).

\subsection{Monopole limit}
Non-integral hyperbolic monopoles were constructed in \cite{nash}.  In the formalism of this report, the monopoles are obtained from (\ref{phi})-(\ref{psi}), with
\[ g(z) = g_M(z) := \exp\left(-\frac{2Bz}{S}\right). \]
$B>1$ is a positive real parameter and the asymptotic norm of the Higgs field of the monopole is equal to $B-1$.

Hyperbolic calorons have a well-defined limit as $\lambda\rightarrow\infty$.  One can show that this limit is gauge equivalent to a hyperbolic monopole, with
\[ B = 1 + \frac{S\pi}{\beta}. \]

\section{Conclusion}
\label{conclusion}
We have investigated the notion of non-integral hyperbolic calorons by constructing explicit examples.  These generalise integral hyperbolic calorons in the same way that non-integral hyperbolic monopoles generalise integral hyperbolic monopoles.  Like Euclidean 1-calorons, hyperbolic 1-calorons have a monopole limit.  The non-integral property also makes it possible to obtain a large period limit, which we have called a hyperbolic instanton.  The zero curvature limits of our hyperbolic calorons are Euclidean calorons.  In the case of hyperbolic monopoles, the zero curvature limit is always a Euclidean monopole \cite{jn97}, but it is not yet known whether a hyperbolic caloron is guaranteed to have a zero curvature limit.

The examples we have given are all rotationally symmetric, but this need not be the case; probably more general examples can be found using the harmonic function ansatz (\ref{hyperbolic hfa1}), (\ref{hyperbolic hfa2}).  It is not clear whether hyperbolic calorons exist with non-trivial asymptotic holonomy, and a more complete account of hyperbolic calorons would address this issue.  Examples of Euclidean calorons with non-trivial holonomy are known \cite{ll,vb98}, but their construction depends on more sophisticated techniques than those presented here.

The study of hyperbolic monopoles, both integral and non-integral, has yielded many interesting results.  For example, hyperbolic monopoles are determined completely by their behaviour on the 2-sphere at infinity \cite{norbury04}.  It would be interesting to see whether this, or other properties of hyperbolic monopoles, have analogues for hyperbolic calorons.

\subsection*{Acknowledgements}
I am grateful to my supervisor, Prof. R. S. Ward, for many useful discussions.  This work is supported by a PPARC studentship.

\bibliographystyle{hplain}
\bibliography{hypcal3}

\begin{thebibliography}{10}

\bibitem{atiyah}
M.~F. Atiyah.
\newblock Magnetic monopoles in hyperbolic spaces.
\newblock In {\em Vector Bundles on Algebraic Varieties}, pages 1--34. Oxford
  University Press, 1087.

\bibitem{chakrabarti85}
A.~Chakrabarti.
\newblock Construction of hyperbolic monopoles.
\newblock {\em J. Math. Phys.}, 27:340--348, 1985.

\bibitem{cf}
E.~Corrigan and D.~B. Fairlie.
\newblock Scalar field theory and exact solutions to a classical $su(2)$-gauge
  theory.
\newblock {\em Phys. Lett.}, 67B:69--71, 1977.

\bibitem{gm}
H.~Garland and M.~K. Murray.
\newblock Why instantons are monopoles.
\newblock {\em Comm. Math. Phys.}, 121:85--90, 1989.

\bibitem{gpy}
D.~J. Gross, R.~D. Pisarski, and L.~G. Yaffe.
\newblock Qcd and instantons at finite temperature.
\newblock {\em Rev. Mod. Phys.}, 53:43, 1978.

\bibitem{hs}
Barry~J. Harrington and Harvey~K. Shepard.
\newblock Periodic euclidean solutions and the finite-temperature yang-mills
  gas.
\newblock {\em Phys. Rev.}, D17(8):2122--2125, 1978.

\bibitem{jt}
Arthur Jaffe and Clifford Taubes.
\newblock {\em Vortices and Monopoles}.
\newblock Birkh\"{a}user, 1980.

\bibitem{vb98}
Thomas~C. Kraan and Pierre van Baal.
\newblock Periodic instantons with non-trivial holonomy.
\newblock {\em Nucl. Phys.}, B533:627--659, 1998, arXiv:hep-th/9805168.

\bibitem{landweber}
Gregory~D. Landweber.
\newblock Singular instantons with $so(3)$ symmetry.
\newblock 2005, arXiv:math.dg/0503611.

\bibitem{ll}
Kimyeong Lee and Changhai Lu.
\newblock $su(2)$ calorons and magnetic monopoles.
\newblock {\em Phys. Rev.}, D15:025011, 1998, arXiv:hep-th/9802108.

\bibitem{manton}
N.~S. Manton.
\newblock Instantons on a line.
\newblock {\em Phys. Lett.}, 76B:111--112, 1978.

\bibitem{nash}
C.~Nash.
\newblock Geometry of hyperbolic monopoles.
\newblock {\em J. Math. Phys}, 27:2160--2164, 1986.

\bibitem{norbury04}
P.~Norbury.
\newblock Asymptotic values of hyperbolic monopoles.
\newblock {\em Journ. of Geom. and Phys.}, 51:13--33, 2004, arXiv:math/9911146.

\bibitem{jn97}
P.~Norbury and S.~Jarvis.
\newblock Zero and infinite curvature limits of hyperbolic monopoles.
\newblock {\em Bull. of the L.M.S.}, 29:737--744, 1997.

\bibitem{rossi}
P.~Rossi.
\newblock Propagation functions in the field of a monopole.
\newblock {\em Nucl. Phys.}, B149:170--188, 1979.

\bibitem{witten}
Edward Witten.
\newblock Some exact multipseudoparticle solutions of classical yang-mills
  theory.
\newblock {\em Phys. Rev. Lett}, 38(3):121--124, 1977.

\end{thebibliography}

\end{document}